\documentclass[twocolumn,pre,showpacs,superscriptaddress]{revtex4}
\usepackage{graphicx, bm}%
\usepackage{color} 
\usepackage{transparent}
\usepackage{psfrag}
\usepackage{dcolumn}
\usepackage{amsmath}
\usepackage{amssymb}
\usepackage{latexsym}
\usepackage{hyperref} 
\usepackage{booktabs}
\usepackage{latexsym}
\begin{document}

\def\bea{\begin{eqnarray}}
\def\eea{\end{eqnarray}}
\def\beq{\begin{equation}}
\def\eeq{\end{equation}}
\def\f{\frac}
\def\k{\kappa}
\def\e{\epsilon}
\def\ve{\varepsilon}
\def\be{\beta}
\def\th{\theta}
\def\t{\tau}
\def\a{\alpha}
\def\J{{\cal J}}
\def\mub{\mu_B}
\def\cDa{{\cal D}[X]}
\def\cDd{{\cal D}[X^\dagger]}
\def\cL{{\cal L}}
\def\cLo{{\cal L}_0}
\def\cLa{{\cal L}_1}
\def\cS{\cal S}
\def\cs{\cal s}
\def\Re{{\rm Re}}
\def\sj{\sum_{j=1}^2}
\def\rk{\rho^{ (k) }}
\def\rek{\rho^{ (1) }}
\def\cek{C^{ (1) }}
\def\rz{\rho^{ (0) }}
\def\D{\Delta}
\def\rt{\rho^{ (2) }}
\def\rtb{\bar \rho^{ (2) }}
\def\trk{\tilde\rho^{ (k) }}
\def\trek{\tilde\rho^{ (1) }}
\def\trz{\tilde\rho^{ (0) }}
\def\trt{\tilde\rho^{ (2) }}
\def\r{\rho}
\def\tD{\tilde {D}}
\def\hm{\hat{m}}
\def\s{\sigma}
\def\kb{k_B}
\def\F{{\cal F}}
\def\la{\langle}
\def\ra{\rangle}
\def\nn{\nonumber}
\def\up{\uparrow}
\def\dn{\downarrow}
\def\S{\Sigma}
\def\dg{\dagger}
\def\d{\delta}
\def\p{\partial}
\def\l{\lambda}
\def\le{\left}
\def\ri{\right}
\def\L{\Lambda}
\def\G{\Gamma}
\def\o{\Omega}
\def\w{\omega}
\def\g{\gamma}
\def\b{\beta}

\def\jv{ {\bf j}}
\def\jr{ {\bf j}_r}
\def\jd{ {\bf j}_d}
\def\noi{\noindent}
\def\a{\alpha}
\def\d{\delta}
\def\p{\partial} 

\def\H{{\bf H}}
\def\He{{\bf H_e}}
\def\h{{\bf h}}
\def\m{{\bf m}}
\def\s{{\bf \sigma}}
\def\hth{h_{\theta}}
\def\vt{\bf \tau}
\def\vxi{\bf \xi}
\def\x{{\bf x}}
\def\vJ{\bf J}
\def\vN{{\bf N}}
\def\Jr{J^{\rm rev}}
\def\Ji{J^{\rm irr}}

\def\la{\langle}
\def\ra{\rangle}
\def\e{\epsilon}
\def\g{\gamma}
\def\G{\Gamma}
\def\break#1{\pagebreak \vspace*{#1}}
\def\hf{\frac{1}{2}}

\title{Stochastic thermodynamics of macrospins with fluctuating amplitude and direction}  
\author{Swarnali Bandopadhyay}
\email{swarnalib@tifrh.res.in}
\affiliation{TIFR Centre for Interdisciplinary Sciences,
21 Brundavan Colony, Narsingi,
Hyderabad 500075, Telengana, India
}
\author{Debasish Chaudhuri}
\email{debc@iith.ac.in}
\affiliation{Indian Institute of Technology Hyderabad,
Yeddumailaram 502205, Telengana, India
}
\author{A. M. Jayannavar}
\email{jayan@iopb.res.in}
\affiliation{Institute of Physics, Sachivalaya Marg, Bhubaneswar 751005, India.
}
\date{\today}

\begin{abstract}
We consider stochastic energy balance and entropy production (EP) in a generalized Langevin dynamics of macrospins, allowing for both amplitude and direction fluctuations, under external magnetic field. 
EP is calculated using Fokker-Planck equation, distinguishing between reversible and irreversible parts of probability currents. 
The system entropy increases due to  irreversible non-equilibrium processes, and reduces as heat dissipates to surrounding environment.  
Using path probability distributions of time-forward trajectories and  conjugate trajectories under time reversal, we obtain 
fluctuation theorems (FT) for total stochastic EP.  We show that the choice of conjugate trajectories is crucial in obtaining entropy like quantities that obey FTs. 

\pacs{05.40.-a, 05.40.Jc, 05.70.-a} 
%\keywords{Fluctuation phenomena statistical physics, Brownian motion, Entropy thermodynamics,Spin polarized transport; Ferromagnetic resonances}
\end{abstract}

\maketitle
\section{Introduction}

Stochastic spin dynamics under magnetic fields and the influence of other spins, plays an important role in understanding magnetic properties of condensed matter systems. With the advent in 
nano-technology, the size of magnetic devices like  magnetic read head and random access memory are being reduced consistently. This makes them vulnerable to thermal 
fluctuations~\cite{Parkin2000,Blanter2000}. Understanding the role of stochasticity in such devices is thus becoming important, even from the perspective of  better control of their performance~\cite{Tserkovnyak2001,Foros2005,Foros2007,Bandopadhyay2011, Covington}. The classical dynamics of a magnetization $\m$ under external field $\H$ is described by the 
Heisenberg motion $\dot \m = \g\, \m \times \H $ where $\g$ denotes the gyromagnetic ratio~\cite{Chaikin1995}. This dynamics, evidently, conserves the amplitude $m = |\m|$. When coupled 
to a heat bath, the dynamics gets stochastic and is often expressed as a stochastic Landau-Lifshitz (sLL) equation~\cite{Kubo1970}
\bea
\dot \m = \g\, \m \times \le[ \le(\H+\h (t) \ri) - \eta' (\m \times \H) \ri].
\label{sLL}
\eea
Kubo and Hashitsume argued for the introduction of dissipation term $-\eta' \m \times(\m \times \H)$ along with the stochastic fluctuation $\h (t)$ to obtain fluctuation-dissipation relation~\cite{Kubo1970}.
Here $\h (t)$ is regarded as a Gaussian white noise with $\la \h (t) \ra =0$, and $\la \h (t) \otimes \h (t') \ra = 2 D_0' {\bf 1} \d(t-t')$ with  ${\bf 1}$ denoting an identity matrix, $D_0' = \eta' \kb T/V$ where $T$ is the temperature, $V$ the physical volume of the macrospin, $\kb$ Boltzmann constant.
The sLL equation was independently derived using the Zwanzig formalism of coupling the spin dynamics with harmonic bath and taking the Markovian limit~\cite{Jayannavar1991}. This equation
describes a stochastic rotational dynamics of magnetization, keeping the magnitude $m$ conserved. 
In Ref.~\cite{Jr1963}, the properties of Fokker-Planck equation corresponding to a related Landau-Lifshitz-Gilbert (LLG) equation were analyzed in detail.
The constant $m$ dynamics is a good approximation for bulk ferromagnets at room temperature, where the transition temperatures for ferromagnetic to 
paramagnetic phase transitions are much larger. % than the room temperature. 
Stochastic fluctuations of $m$ occur within a ferromagnetic domain due to exchange interaction, with enhanced effect near the
transition temperature~\cite{Chaikin1995}. 
The fluctuations in $m$ becomes dominant in bulk ferromagnets only at high temperatures. 
 On the other hand, due to enhanced relative fluctuations in small ferromagnetic domains, e.g., in a macrospin, the 
{ transition temperatures} get largely suppressed with reduction of system size~\cite{Velasquez2011,Bertoldi2012}, enhancing the fluctuations in $m$ even at room temperature.
Recently, a generalized Langevin spin dynamics has been proposed that captures longitudinal fluctuations in the spin magnitude, 
as well as the stochastic rotation dynamics of its orientation~\cite{Ma2012}.  
In this paper, we present stochastic thermodynamics of a macrospin system, deriving  stochastic energy balance relation
and fluctuation theorems for probability of  entropy production.

During the last two decades, a theoretical description of stochastic thermodynamics has been developed to describe non-equilibrium small systems having
enhanced relative fluctuations, using stochastic counterparts of thermodynamic variables like energy, work, 
entropy etc.~\cite{Sekimoto1998,Lebowitz1999,Crooks1999,Jarzynski1997,Narayan2004,Imparato2006,Kurchan2007,Saha2009}. 
While the possibility of second law violating stochastic trajectories was recognized long back~\cite{Chandrasekhar1943}, it took several decades before it was shown that 
probabilities of such trajectories  in steady state $P(-\D s_t)$, with $-\D s_t$ denoting negative entropy production,  are exponentially suppressed with respect to the positive entropy producing ones
via the relation $P(\D  s_t)/P(-\D s_t) = \exp(\D s_t/\kb)$~\cite{Evans1993,Evans1994,Gallavotti1995}. This relation is known as the detailed fluctuation theorem. This, and a related integral fluctuation theorem 
$\la \exp(-\D s_t/\kb) \ra =1$, which is equivalent to the Jarzynski equality for non-equilibrium transformations from an initial equilibrium state to a final state that eventually reaches equilibrium,  have been derived~\cite{Bustamante2005,Jarzynski2011,Seifert2012, Jarzynski1997,Lebowitz1999,Crooks1999,Seifert2005}. 
These theorems were verified using experiments on colloids~\cite{Wang2002,Blickle2006,Speck2007},  granular matter~\cite{Feitosa2004,Joubaud2012}, and used to obtain the free energy landscape of RNA~\cite{Liphardt2002,Collin2005}, and torque produced by F1-ATPase motor proteins~\cite{Hayashi2010}.

In the following section, we present the generalized Langevin dynamics of macrospins (GLDM). We discuss its motivation, and corresponding projected dynamics along the longitudinal and transverse directions. Next, we study its stochastic thermodynamics, first deriving the stochastic energy balance, and then entropy production using Fokker-Planck equation and ratio of time-forward and conjugate path-probabilities.
Our analysis shows that it is possible to obtain fluctuation theorems for entropy like quantities, each of which emerges out of a specific way of choosing conjugate trajectories. The time-reversed trajectories 
give fluctuation theorems in terms of EP in reservoir given by the dissipated heat, which is consistent with the results of Fokker-Planck equation. Another possible choice of conjugate trajectories leads to an entropy like quantity that also involves gyroscopic work done due to magnetic field induced spin torque. This quantity also obeys both detailed and integral fluctuation theorems. 
We present discussions interpreting our results.   Finally, we conclude by presenting a summary.

\section{Model}
Consider a macrospin having magnetization $\m$, and volume $V$.   The GLDM for the macrospin in presence of a time-dependent external magnetic field $\H(t)$ can be written as~\cite{Ma2012}
\begin{equation}
\dot \m=  \g\, \le[ \m \,\times\,\H (t) -  \eta \f{\p g}{\p \m} +\h (t)\ri]\,,
\label{gldm1}
\end{equation}
where $\dot \m \equiv d\m/dt$.
The Langevin heat bath is characterized by the dissipation coefficient $\eta$, and the Gaussian white noise $\h$ the  components of which obey $\la \h (t) \ra =0$, 
$\la \h (t) \otimes \h(t')\ra = 2  D_0 {\bf 1} \d(t-t')$ with $D_0=\eta \kb T/V$. 
In the above equation $\m \times \H$ denotes a non-conservative spin-torque.
The  energy density is given by  
\bea
g &=&  (f_L - \m.\H (t)) \nn\\
{\rm where,}~ f_L &=& -\f{a}{2} m^2 + \f{b}{4} m^4,
\label{free}
%\end{align}
\eea
is the Landau free energy density,
and the effective magnetic field $\H^{\rm eff} = -\p g/\p \m = \H^{\rm int} + \H (t)$, with $\H^{\rm int} = (a - b m^2)\m$ being the mean field contribution due to collective spin alignment. 
$g$ can be expressed as $g=-\m.\H^{\rm eff}$.
$f_L$ denotes the Landau free energy density having two equivalent minima at $\m = \pm \sqrt{a/b}\, \hm$, 
with $a = a_0 (T_c-T) > 0$ in the ferromagnetic phase, 
where $T_c$ is the  transition temperature, and $T$ is the temperature of the system~\cite{Chaikin1995}. With reduction of system size, $T_c$ decreases. 
It was shown for three dimensional Ising clusters with total number of spins $N$, the transition temperature decreases with reduction in 
macrospin size $N$ as $T_c \sim T_c^\infty(1- 1/N^\phi)$ where $\phi \approx 1/3$, and $T_c^\infty$ denotes the transition temperature of thermodynamically large system~\cite{Bertoldi2012}.
Thus for small enough size of a macrospin, $T_c$ approaches $T$ from above, thereby increasing fluctuations in $m$.
The term $-\m.\H$ in $g$ is due to external magnetic field $\H$,
 and shifts the global minimum towards positive $\hm$.
Thus the GLDM may be expressed as
\bea
\dot \m =   \le[ \m \,\times\,\H (t)  + \eta \H^{\rm eff} (t)+ \h (t) \ri ],
\label{gldm2}
\eea
absorbing $\g$ into the definition of time, $t \to \g t$.

Eq.~(\ref{gldm1}) may be motivated by drawing parallel to Langevin equations of motion of driven diffusing particles~\cite{Ma2012}. 
Note that such a particle in one dimension obeys
$\dot v = f(t) - \eta v + {\xi}$,
where $v$ denotes the particle velocity, $f(t)$ a time-dependent external force. The viscous dissipation $ - \eta v $ and the Gaussian white noise 
$\xi$ are forces due to coupling to the heat bath, with $\la \xi(t) \ra =0$, and $\la \xi(t) \xi (t') \ra = 2 \eta \kb T \d(t-t')$.
The viscous dissipation $-\eta v$ may be rewritten as $-\eta  \p_{v} {\mathcal H}$ given that 
the velocity dependence of the Hamiltonian ${\mathcal H}$ is ${v^2}/{2}$.
Using this as a guiding principle, one may replace $v$ by the magnetic moment $\m$, as both are
odd parity variables under time reversal. Similarly, the force $f$ may be replaced by the external torque due to the magnetic field $\m \times \H$.
Replacing Hamiltonian $\mathcal{H}$ by  energy density $g=-\m.\H $ for a single spin, and $\p_v \mathcal{H}$ by $\p g/\p \m = -\H$ we obtain 
the GLDM for a single spin $\dot \m=   \le[ \m \,\times\,\H   + \eta \H + \h (t) \ri ]$. 
Note that in this equation, the term $\eta \H +\h$ denotes the force and torque due to the heat bath~\cite{Ma2012}.
Extending this argument to a macrospin containing large number of spins,
one obtains Eq.(\ref{gldm2}) by using $g = -\m.\H + f_L$. In Eq.(\ref{gldm2}), the term $\eta \H^{\rm eff} +\h$ denotes the force and torque on the macrospin due to the heat bath.
Throughout this paper, we use Stratonovich convention while interpreting stochastic differential equations.

It is possible to separate the longitudinal and transverse dynamics of the macrospin $\m$. 
Taking longitudinal projection, i.e., projecting Eq.~(\ref{gldm1}) along $\hm = \m/m$ we obtain the dynamics for the spin amplitude
\beq
\dot m= \le [ \eta H_\parallel (t)+ h_\parallel (t) +\eta (a m - b m^3)\ri] \, ,\label{long}
\eeq
where $H_\parallel (t) = \hm.\H (t)$  and $h_\parallel = \hm.\h$, with $\la h_\parallel(t) \ra = 0$, and 
$\la h_\parallel(t) h_\parallel(t') \ra = 2D_0  \d(t-t')$. 
Clearly, in this equation, $\eta (H_\parallel + a m - b m^3) + h_\parallel$ is the longitudinal force  due to the Langevin heat bath.
The corresponding Fokker-Planck equation is $\p_t P(m,t) = -\p_m j$ where $j = -D_0  \p_m P + \eta [a m - b m^3 + H_\parallel]$. 
For a time-independent magnetic field, setting the dissipative current $j=0$, one obtains the detailed balanced equilibrium distribution,
$P_{\rm eq}(m)=P_0 \exp(-g_\parallel V/\kb T)$ with the energy density 
$g_\parallel = f_L  -H_\parallel m$. 

Subtracting the longitudinal dynamics Eq.(\ref{long}) from Eq.(\ref{gldm2}), one obtains 
\bea
\dot \m_\perp = \m \times \H(t) + \eta \H_\perp(t) + \h_\perp (t), 
\label{m_perp}
\eea
where 
$\m_\perp = \m - \hm m$, $\H_\perp = \H - \hm H_\parallel = -\hm\times(\hm\times\H)$, and $\h_\perp = \h - \hm h_\parallel$. 
It can be shown that $\h_\perp$ and $\tilde \h_\perp =  \hm\times\h$ obeys the same statistics : $\la \h_\perp \ra =0 = \la \tilde \h_\perp \ra$, 
$ \la  \h_\perp (t)  \otimes \h_\perp (t') \ra  = ( {\bf 1}  - \hm \otimes \hm ) \d(t-t') =  \la \tilde \h_\perp (t) \otimes \tilde \h_\perp (t') \ra$~\cite{Ma2012}.
Thus one can replace $\h_\perp$ by $\tilde \h_\perp$ in Eq.(\ref{m_perp}). The resultant equation can be expressed as
\beq
\dot \m_\perp = \m \,\times (\H (t) + \h' (t)\,) -\eta' \m \times(\m \times\H (t)\, ),\label{sll}
\eeq
where $\h' = \h/m$, and $\eta'= \eta/m^2$. Note that for constant magnitude $m$, $\dot \m_\perp = \dot \m$, and Eq.(\ref{sll}) is then same as the sLL equation Eq.(\ref{sLL}). 
The term $-\eta' \m \times(\m \times\H)$ denotes the Gilbert damping, and  $\m \,\times \h'$ denotes the stochastic part of the total torque imparted due to the Langevin heat bath.

\section{Results and Discussions}
\subsection{Energy conservation}
The macrospin undergoes a relaxation dynamics in the Langevin heat bath, settling into an average unidirectional precession around the field $\H$. Unlike sLL equation which describes motion of spin constrained to have a constant magnitude, here amplitude $m=|\m|$ is not conserved, and obeys the above mentioned distribution $P_{\rm eq}(m)$. The rate of change of energy density $\dot g$ of the macrospin is given by $\dot g=-\dot \m.\H^{\rm eff}\,-\m.\dot \H $, with $\H^{\rm eff}=\H + H_{\rm int}$ and $\H_{\rm int} = -\p f_L/\p \m$. 
Note that $\H$ and $\H_{\rm int}$ shares the same symmetry under time reversal, as does $\H$ and $\m$. 
Substituting Eq.~(\ref{gldm1}) in the expression of $\dot g$ we obtain stochastic energy balance,
\bea
 \dot g &=& \dot q+\dot w\,,\label{law1}\\
 \mbox{where}\,\,\,\dot q &=& - \dot \m.\H^{\rm eff} \,\label{heat}\\
 \mbox{and}\,\,\,\dot w &=& -\m.\dot\H\,\,.
\label{work}
\eea 
We used the sign convention that stochastic heat and work done are positive 
if they increase the energy of the system. 
Note that the stochastic energy balance presented above, is a relation between energy density, work density and heat absorption per unit volume.
Here $\dot q$ and $\dot w$ represent the rate of heat absorbed by the system and the rate of work done on the system respectively. 
Note that for driven diffusive particles, heat absorbed by the system is given by $v(-\eta v +\xi)$, and we motivated the GLDM equation
by replacing $v$ by $-\H$, and $\xi$ by $\h$. Thus it is only natural to identify $-(\eta H^2 + \h.\H)$ as heat absorbed by a single spin. For
a system of spins present in the macrospin, $\H$ has to be replaced by $\H^{\rm eff}$. Thus $\dot q = - [\eta\,(H^{\rm eff})^2\,+\h.\H^{\rm eff}] = - \dot \m.\H^{\rm eff}$. 
This suggests a stochastic version of 
Clausius entropy production in the heat bath in the form $-\dot q/T$. In the following, we present a careful analysis of entropy production.

\subsection{Entropy production and heat dissipation}
\label{FPeqn}
With $P(\m,t)$ denoting the probability of finding a spin in the state $\m$ at time $t$, the non-equilibrium Gibbs entropy $S = -\kb \int d\m\, P \ln P(\m,t)$
suggests a definition of time dependent stochastic entropy of the system $s(t) = -\kb \ln P(\m,t)$
where $S = \la s \ra$ denotes the ensemble average of stochastic entropy~\cite{Seifert2005}.
Note that, here and in the rest of the paper, whenever we mention entropy, energy or work done, it actually means the quantity {\em per unit volume}. 
To achieve this in the following, we replace diffusivity of magnetization $D_0 $ by $D = \eta \kb T$.  

The rate of change in stochastic entropy  is given by
\beq
\f{\dot s}{\kb} = -\f{\p_t P}{P} - \f{\p_{\m} P}{P}. \dot \m \,.
\label{sdot1}
\eeq
where the probability density $P(\m,t)$ obeys the Fokker-Planck equation
 $\p_t P = - \p_{\m}. \vJ$ with  $\p_{\m} \equiv (\p_{m_x}, \p_{m_y},\p_{m_z})$  
and probability flux $\vJ$. 
Note that under time reversal $t$, $\m$ and $\H$ change sign. Thus $\vJ = {\bf \Jr} + {\bf \Ji}$, where ${\bf \Jr}$ is the reversible current that does not change sign under time reversal, 
and ${\bf \Ji}$  is the irreversible current that changes sign~\cite{Spinney2012}. 
The $i$-th component of these currents are given by 
\bea
\Jr_i &=& N_i P \nn\\
\Ji_i &=&  \eta\,H^{\rm eff}_i  P - D \p_{m_i} P\,,
\label{current}
\eea
where, the component of spin-torque due to external magnetic field $N_i = (\m \times \H)_i$. 
Using Eq.~(\ref{current}) in Eq.~(\ref{sdot1}) to replace $\p_{\m} P$, one can express the rate of change in stochastic entropy as
\bea
\f{\dot s}{\kb} &=& -\f{\p_t P}{P} +\f{\Ji_i\dot m_i}{PD} - \f{\eta}{D} H^{\rm eff}_i  \dot m_i \nn \\
&=& -\f{\p_t P}{P} +\f{\Ji_i\dot m_i}{PD} - \f{1}{\kb T} \dot q \, .
\label{sdot}
\eea

At this stage, let us perform a two step averaging, (i)~over trajectories, and (ii)~over phase space by integrating over all $\m$ with probability $P(\m,t)$. 
The trajectory average of the components of magnetization dynamics depends on  both reversible and irreversible parts of probability flux,
$\la \dot m_i |\, \m, t\ra = J_i/P = (\Jr_i + \Ji_i)/P = N_i + \Ji_i/P$~\cite{Seifert2005}. 
Thus after the trajectory average one can replace $(\Ji_i \dot m_i) / (P D)$ by $[(\Ji_i N_i )/(P D) + (\Ji_i)^2/ (P^2 D)]$.
Now let us perform averaging over the probability density $P(\m,t)$ by multiplying Eq.(\ref{sdot}) throughout by $P(\m,t)$ and performing integration over $\m$. 
The conservation of probability $\int d\m P =1$ leads to $\int d\m\, \p_t P =0$. Thus one obtains the final average
\bea
\f{\dot S}{\kb} = \f{\la \dot s \ra}{\kb} = \f{1}{D} \int d\m\, \f{(\Ji_i)^2}{P} + \f{1}{D} \int d\m\, \Ji_i N_i - \f{\la \dot q \ra}{\kb T}. \nn
\eea
Now using the expression in Eq.(\ref{current}) one can show that the second term in the above equation $\int d\m\, \Ji_i N_i =0$. 
This term vanishes, as (i)~$H^{\rm eff}_i N_i = 0$ due to vector identities  $\H.(\m \times \H) = 0$ and $\m.(\m \times \H) = 0$,
(ii)~$\int d\m\, N_i  \p_{m_i} P  = 0$ using integration by parts. 
Thus
\bea
{\dot S} = {\la \dot s \ra} = \f{1}{\eta T} \int d\m\, \f{(\Ji_i)^2}{P}  - \f{\la \dot q \ra}{T} \equiv \Pi -\Phi, 
\label{dotS}
\eea
where $\Pi = \f{1}{\eta T} \int d\m\, \f{(\Ji_i)^2}{P} $ is the EP in the system due to irreversible processes quantified by $\Ji_i$, and 
$\Phi = \la \dot q \ra/T$ is the entropy flux to the reservoir due to average heat loss. At this stage, it is interesting to note that as the system gets isolated from the heat bath, i.e., $\eta \to 0$,
$\Pi \sim \eta \to 0$, a result expected for EP in an isolated system. 
The total EP in the combined system and reservoir
obeys the second law of thermodynamics, $\dot S_t = \dot S + \Phi = \f{1}{\eta T} \int d\m\, \f{(\Ji_i)^2}{P} \geq 0$, where the equality denotes equilibrium with $\Ji_i = 0$. 
At steady state, $\Phi = \Pi = \la \dot q \ra /T$, and average change in energy $\la \dot g \ra =0$ leads to $\la \dot q \ra = -\la \dot w \ra = \la \m \cdot \dot\H \ra$. Thus one
can express the average entropy flux as $\Phi = \la \m \cdot \dot\H \ra/T$. The steady state EP in the reservoir is due to non-equilibrium processes driven by time-dependent external field $\H(t)$.

The above discussion shows that the stochastic EP in the reservoir is
\bea
\dot s_r = -\f{\dot q}{T}.
\label{srdot}
\eea
{%\color{red}
This quantity can be both positive or negative. Oono and Paniconi~\cite{Oono1998} introduced a concept of housekeeping heat, which is the heat dissipated to keep the system at 
non-equilibrium steady state. As we have seen above, at steady state, the average heat dissipated  is equal to the mean work done by the system $\la -\m \cdot \dot\H\ra$. Thus the expression of 
stochastic housekeeping heat generation $\dot q_h = - \m \cdot \dot\H$. This gives the rate of excess heat generation $\dot q_e = \dot q - \dot q_h = \m\cdot \dot \H - \dot \m \cdot \H^{\rm eff}$. 
If one changes the magnetic field from some initial value to a final value, the average excess heat generation
remains non-zero transiently before the system relaxes from one steady state to another.
}

\subsection{Equilibrium detailed balance}
The steady state condition is given by $\p_{m_i} [\Jr_i + \Ji_i] = 0$. At equilibrium, the dissipative current must vanish, $\Ji_i=0$. This leads to the condition 
$dP/P = \be H^{\rm eff}_i  d m_i$. The equation can be integrated for a time-independent external field $\H$ to give 
$$P=P_0 \exp[-\be \{ f_L -  \m\cdot\H\}],$$ 
where $f_L = -(a/2) m^2 + (b/4) m^4$.
Using the relation $\Ji_i=0$ in the steady state condition one obtains $\p_{m_i} \Jr_i =0$, which is readily obeyed. These two relations,  $\Ji_i=0$ and $\p_{m_i} \Jr_i =0$ define the equilibrium detailed balance 
condition. A time-dependent magnetic field brings the system out of equilibrium, and allows EP. 

\subsection{Entropy production using path probabilities: Fluctuation theorems}
\label{EPtrajectory}
EP along  stochastic trajectories of a non-equilibrium system may also be estimated by using the inequality of probabilities of time-forward trajectories, and  conjugate trajectories under suitably time-reversed protocol. We consider the time evolution of a macrospin from $t=0$ to $\t_0$ through a path $X = [\m(t),\H(t)]$ where $\H(t)$ acts as control parameter, the functional form of which gives a specific protocol. 
Let us divide the path into $i=1,2,\dots,N$ segments, each of time-interval $\d t$ such that $N \d t=\t_0$. The transition probability $p_i^+ (\m', t+\d t |\m,t)$ on $i$-th infinitesimal segment is governed by the Gaussian random noise $\h_i$ at $i$-th instant obeying probability distribution $P(\h_i) = (\d t/4\pi D)^{1/2} \exp(-\d t\, \h_i^2/4 D)$. Denoting Eq.(\ref{gldm1}) as $\dot \m= \Phi(\m(t), \H(t))$, the transition probability on $i$-th segment
$p_i^+ = \J^+_i  \int d \h_i P(\h_i) \d(\dot m_i -\Phi_i)$, where
the Jacobian of transformation at $i$-th instant of time $\J^+_i={\rm det}[ \, (\p \h/\p \m)_i\, ]$. 
Using Stratonovich discretization, one can show 
\beq
\J^+_i  = \f{1}{\d t} \le[ 1 - \f{\d t}{2} \f{\p {\bm {\mathcal F(\m_i)}}}{\p \m_i}  \ri]
\label{jacobian1}
\eeq
where ${\bm {\mathcal F(\m_i)}} = (\m \times \H)_i + \eta[(a-b m^2)\m]_i + \eta \H_i$. Note that $\p  (\m \times \H)_i / \p \m_i =0 $, and $\p \H_i /\p \m_i =0 $. Thus the operative part of ${\bm {\mathcal F(\m_i)}}$ in the above relation is the effective field contribution $\H^{\rm int}_i =  \eta[(a-b m^2)\m]_i $.  Eq.(\ref{jacobian1}) can be expressed as
\bea
\J^+_i  = \f{1}{\d t} \le[ 1 - \f{\d t}{2} \f{\p \H^{\rm int}_i {(\m_i)}}{\p \m_i}  \ri]
\label{jacobian}
\eea
The probability of a complete trajectory is ${\cal P}_+ = \prod_{i=1}^N p_i^+$.

%%%%%%%%%%%%%
Similarly, the conjugate trajectory under time-reversal may be discretized, and the probability of such complete trajectories may be expressed as ${\cal P}_- = \prod_{i=1}^N p_i^-$.
There exists various possibilities to choose conjugate  trajectories under time-reversed protocol~\cite{Seifert2012,Speck2008,Bandopadhyay2015}. 
The conjugate trajectory must be carefully  chosen so that the ratio ${\cal P}_+ /{\cal P}_-$ serves as a measure of irreversibility of the process, and as a result characterizes EP
in the surrounding environment. 

Under time-reversal, $\H$ and $\m$ changes sign simultaneously. The corresponding conjugate trajectory is denoted by  $X^\dagger = [-\m(\t_0 - t), -\H(\t_0 -t)]$.
This is similar to requirement of reversal of external flow direction in Ref.~\cite{Speck2008}, under time reversal.
 %%%%%%%%%%%%%%%
The probability of time-reversed trajectory   ${\cal P}_- = \prod_{i=1}^N p_i^-$, where 
$p_i^- =  \J^-_i \int d \h_i P(\h_i)\, \d(\dot \m_i -\Phi_i(\t_0 -t)\,)$. 
It is easy to see from Eq.(\ref{jacobian}) that $\J^+_i = \J^-_i$.
After some algebra, one obtains the ratio of the two probabilities of forward and reverse paths 
$\f{{\cal P}_+}{{\cal P}_- } = \exp(\D s_r/\kb)$, where 
\bea
\f{\D s_r}{\kb}   
=\f{\eta}{D} \int_0^{\t_0} dt\, \H^{\rm eff}.\dot \m= - \f{\D q}{\kb T}.
\label{dsr}                                        
\eea
Note that the expression of $\D s_r$ presented above agrees with the EP given in Eq.(\ref{srdot}).
Let us now assume that $s_{0}$ and $s_{\ell}$ are stochastic entropies of the system corresponding to its initial and final steady states respectively. So, the change in stochastic system entropy
 is given by $\D s = s_{\ell} - s_{0} = \kb \ln (P_{0}/P_{\ell})$ where $P_{0}(\m_{0},\H_{0})$ and $P_{\ell}(\m_{\ell},\H_{\ell})$ are distribution functions of these micro-states. 
 
 %%%%%%%%%%%%%%%%%%
 As we have shown above,  the change in reservoir entropy depends on the trajectory and is given by $\D s_r = \kb \ln ({\cal P}_+ /{\cal P}_-)$. Thus the total entropy change 
 \beq
\D s_t = \kb \ln \le( \f{P_0 {\cal P}_+ }{ P_\ell {\cal P}_-} \ri) = \D s + \D s_r.
\label{stot}
\eeq 
This immediately implies an integral fluctuation theorem (IFT) $\la e^{-\D s_t/\kb} \ra = 1$~\cite{Seifert2012}. Note that in deriving IFT, $\sum_X \equiv \sum_X^\dagger$ is used, as the Jacobian of transformation from time-forward path $X$ to time-reversed path $X^\dagger$ is unity~\cite{Spinney2012}. Further, in a steady state, the total entropy change along a time-forward path $\D s^f_t$ is equal and opposite to that along the time-reversed path, $\D s^r_t(X^\dagger) = -\D s^f_t(X)$. 
Using this, and Eq.(\ref{stot}) one obtains the following detailed fluctuation theorem (DFT)~\cite{Kurchan2007, Crooks1999} for probability distribution of EP $\r(\D s_t)$   as
\bea
\r(\D s_t) &=& e^{\D s_t/\kb} \rho(-\D s_t).
\label{dft}
\eea

Using the definition $\D s_r = -\D q/T $ the IFT $\la \exp(-\D s_t/\kb)\ra =1$ can be expressed as 
\beq
\la \exp(\be \D q - \D s/\kb)\ra = 1. 
 \eeq
This is equivalent to Jarzynski relation, for transformations between non-equilibrium steady states~\cite{Jarzynski1997,Hatano2001}.
Due to Jensen inequality, this implies $T\la \D s \ra \geq \la \D q \ra $. For an infinitesimally slow variation of $\H(t)$, the equality holds, i.e., 
the steady state change in system entropy can be evaluated in terms of $\la \D q \ra \approx \la -\m\cdot \D \H \ra$.
For a time-independent external field, one reaches an equilibrium steady state with $\la \D q \ra =0$, and  $\la \D s \ra = 0$. 

%%%%%%%%%%%%%%%%%%%%%%%
\subsection{Other possibilities of conjugate trajectories}
\label{others}
Let us now consider, three other possibilities of choosing conjugate trajectories, such that one obtains entropy like quantities that obey DFT~\cite{Spinney2012a}. 
First, assume conjugate trajectories such that time forward protocol of $\H(t)$ traces back itself under time reversal. The corresponding conjugate trajectory is denoted by
$X^\dagger = [\m(\t_0-t), \H(\t_0-t)]$ where $\m$ and $\H$ do not change sign. The probability of such conjugate trajectories is denoted by ${\cal P}_-^{(1)}$. 
Then the ratio of probabilities of time-forward and conjugate trajectories is ${\cal P}_+ / {\cal P}_-^{(1)} = \exp(\D s_r^{(1)}/\kb)$ where 
\bea
\f{\D s_r^{(1)}}{\kb} &=&  \f{\eta}{D} \int_0^{\t_0} dt\, \H^{\rm eff} \cdot \dot \m + \f{1}{D}  \int_0^{\t_0} dt \, \vN\cdot \dot\m \nn\\
 &=& \f{1}{\kb} [ \D s_r + \D s_{\rm gyro} ] \, ,
\label{dsr1}                                        
\eea
where $\D s_{\rm gyro} =  \D w_{\rm gyro}/T$ 
with $\D w_{\rm gyro} = (1/\eta) \int dt \vN \cdot \dot \m$ being the gyroscopic work done on the system due to spin torque.
One obtains the DFT 
$$\r(\D s_t^{(1)}) = e^{\D s_t^{(1)}/\kb} \rho(-\D s_t^{(1)})$$ 
where
$\D s_t^{(1)} = \D s + \D s_r^{(1)}$. Numerical simulation of macrospins with constant amplitude $m$ has been used in Ref.~\cite{Bandopadhyay2015} to obtain the 
probability distribution $\r(\D s_t^{(1)})$, which obeys DFT.
This form of DFT may be interpreted as follows. One can define $\D \tilde s_t = \D s - \D q/T$, and rewrite the DFT as,
\bea
\f {\r(\D \tilde s_t, \D w_{\rm gyro} )}{\rho(-\D \tilde s_t, -\D w_{\rm gyro})} &=& e^{\f{1}{\kb} (\D \tilde s_t +  \f{\D w_{\rm gyro}}{T})}. \nn
\eea
In a steady state, ignoring $\D s$ with respect to $\D q/T$, this relation leads to
\bea
\f{\r(-\D q, \D w_{\rm gyro})}{\r(\D q, -\D w_{\rm gyro})} = e^{-\b(\D q - \D w_{\rm gyro})} .
\label{dft2}
\eea
This equality is closely related to the fluctuation theorem for
heat engines~\cite{Sinitsyn2011,Lahiri2012a,Rana2014}, and was used in Ref.~\cite{Utsumi2015} in the context of an isothermal engine 
absorbing heat $\D q$ and performing work $\D w_{\rm gyro}$ via spin torque.

The Jacobian of transformation from time forward trajectory $X$
and the conjugate trajectory $X^\dagger$ is unity. This leads to the IFT  $\la e^{-\D s_t^{(1)}/\kb}\ra =1$, which by Jensen's inequality gives $\la \D s_t^{(1)} \ra \geq 0$, a result equivalent to 
the second law of thermodynamics. 
The IFT obtained from Eq.(\ref{dft2}) has the form $\la e^{-\b(\D q - \D w_{\rm gyro})}\ra=1$, which after Jensen's inequality gives $\la \D w_{\rm gyro} \ra/ \la \D q\ra \leq 1$, meaning average work does
not exceed average heat. 
Note that the torque $\vN$ is associated with the reversible part of probability current ${\bf \Jr} = \vN P$, and thus does not contribute to heat flux. However, it still contributes towards an entropy like 
term $\D s_r^{(1)}$ that gives total entropy $\D s_t^{(1)}$  obeying DFT and IFT.

Next we assume that $\m$ alone changes sign along the conjugate trajectories so that they are described by $X^\dagger = [-\m(\t_0-t), \H(\t_0-t)]$. We denote the path probabilities along such conjugate trajectories by ${\cal P}_-^{(2)}$. Then the ratio ${\cal P}_+ / {\cal P}_-^{(2)} = \exp(\D s_r^{(2)}/\kb)$ where
\beq
\f{\D s_r^{(2)}}{\kb} = \f{1}{ D}\,\int_0^{\t_0} dt\, \vN. \dot{\m} \equiv \f{\D s_{\rm gyro}}{\kb}.
\label{dsr2}
\eeq
Again, $\D s_t^{(2)} = \D s + \D s_r^{(2)}$ obeys the DFT. However, the Jacobian of transformation from $X$ to $X^\dagger$ is not unity, and the IFT is not obeyed by this quantity. This is expected,
as $\D s_r^{(2)}$ depends only on $\vN$, which is associated with reversible probability current, and should not give rise to second law like inequality. 

The third alternative is to consider conjugate trajectories in which $\H$ alone changes sign, i.e., $X^\dagger = [\m(\t_0-t), -\H(\t_0-t)]$. Denoting the probability of conjugate trajectory ${\cal P}_-^{(3)}$,
one obtains ${\cal P}_+ / {\cal P}_-^{(3)} = 1$, i.e., the corresponding stochastic EP in the reservoir $\D s_r^{(3)}=0$. 

The EP in reservoir associated with dissipated heat $\D s_r = \D s_r^{(1)} - \D s_r^{(2)}$. Note that the amplitude of magnetization can be approximated to be constant, for samples with Curie temperature much larger than room temperature. In such cases, the stochastic Langevin dynamics can be described as diffusion of a particle under suitable torque due to external field~\cite{Bandopadhyay2015}. In the spherical polar coordinates, macrospin orientation $(\th, \phi)$ may be treated as even functions under time reversal. As a result one obtains an expression of entropy, which is equivalent to $\D s_r^{(1)}$  involving a gyroscopic term $\D s_{\rm gyro}$. The probability distribution of total EP $\r(\D s_t^{(1)} )$ has been shown to obey DFT. Of course, even within that restricted dynamics, if one considers $X^\dagger = [-\m(\t_0-t), -\H(\t_0-t)]$ as the conjugate trajectory, one obtains EP in the reservoir $\D s_r = - \D q/T$, as is shown in the appendix of Ref.~\cite{Bandopadhyay2015}.  

Among all possible prescriptions for constructing stochastic trajectories, the definition of 
$\D s_r$ in Eq.(\ref{dsr}) obtained by tracing back the time-reversed trajectory directly utilizing reverse protocol of $\H(t)$, such that, $X^\dagger = [-\m(\t_0-t), -\H(\t_0-t)]$ leads to 
the expression $\dot s_r$ in Eq.(\ref{srdot}) obtained from Fokker-Planck equation. The ratio of probabilities of time forward, and time reversed trajectories gives unity in presence of 
time reversal symmetry. Thus any other value of this ratio gives a measure of breaking of time-reversal symmetry, and thus the EP.
Note that the derivation of $\dot s_r$ in Eq.(\ref{srdot}) depends only on the dynamics, not on any particular definition of conjugate trajectory. Such definitions were used as mathematical construct
to derive fluctuation theorems.

%%%%%%%%%%%%%%%%%%%%%%%%%%%%%%%
\section{Summary}
We studied stochastic thermodynamics for a macrospin of fluctuating amplitude and direction of magnetization subjected to external magnetic field. We considered a generalized Langevin dynamics of macrospins, taking into account  (i)\,a stochastic rotational dynamics of the magnetization and (ii)\,its longitudinal fluctuations, (iii)\, a mean field approximation of the interaction between spins within the macrospin cluster, and an external magnetic field.  We obtained several possible fluctuation theorems for entropy-like quantities found from using different choices of conjugate trajectories under time reversal.  Only one of the possible choices gave $\D s_r=-\D q/T$, the entropy production (EP) in the reservoir due to dissipated heat $-\D q$, that agrees with the expression one obtains from Fokker-Planck equation. 
A second entropy like quantity $\D s_t^{(1)} = \D s -\D q/T + \D w_{\rm gyro}/T$, where $\D w_{\rm gyro}$ is the rotational work done on the macrospin due to magnetic field induced spin torque, also obeys fluctuation theorems. 
The heat dissipation and gyroscopic work done, can be measured separately in experiments on macrospins, and our predictions regarding fluctuation theorems can be tested.

\acknowledgments
DC thanks Madan Rao for stimulating discussions, and Simons Centre at NCBS, Bangalore for hospitality while writing the paper. AMJ thanks DST, India for financial support. 
We thank the anonymous referees for illuminating comments.

\bibliographystyle{prsty}

\end{document}